%% file: main.tex
\documentclass[a4paper,11pt]{article}
\usepackage[utf8]{inputenc}
\usepackage{xcolor}
\usepackage{hyperref}
\definecolor{linkcolor}{HTML}{799B03}
\definecolor{urlcolor}{HTML}{799B03}
\hypersetup{pdfstartview=FitH,linkcolor=linkcolor,urlcolor=urlcolor,colorlinks=true}
\usepackage[english]{babel}
\usepackage{amsmath, amsthm, amsfonts, amssymb,amscd}

\usepackage[all]{xy}
\ifx\pdfoutput\undefined
\usepackage{graphicx}
\else
\usepackage[pdftex]{graphicx}
\fi
\usepackage{import}
\usepackage{xifthen}
\usepackage{pdfpages}
\usepackage{float}
\usepackage{transparent}
\usepackage{multirow}
\usepackage{cellspace}
\usepackage{colortbl}
\usepackage{bbm}
\usepackage{mathrsfs}
\usepackage[backend=biber, citestyle=numeric-comp, sorting=none]{biblatex}
\usepackage{csquotes}
\graphicspath{{pic/}}
\def\[{\begin{equation}}
\def\]{\end{equation}}
\def\X{\mathcal{X}}
\def\Y{\mathcal{Y}}
\def\Z{\mathcal{Z}}
\DeclareUnicodeCharacter{0301}{`}

\begin{document}

\begin{center}
 {\LARGE \bf Perturbations in Horndeski theory above anisotropic cosmological background}

 \vspace{20pt}

 S. Mironov$^{a,b,c}$\footnote{sa.mironov\_1@physics.msu.ru},
 A. Shtennikova$^{a,b,d}$\footnote{shtennikova@inr.ru}

 \renewcommand*{\thefootnote}{\arabic{footnote}}
 \vspace{15pt}

 $^a$\textit{Institute for Nuclear Research of the Russian Academy of Sciences,\\
  60th October Anniversary Prospect, 7a, 117312 Moscow, Russia}\\
 \vspace{5pt}
 $^b$\textit{Institute for Theoretical and Mathematical Physics,\\ Lomonosov Moscow State University, Moscow
  119991, Russia}\\
 $^c$\textit{NRC “Kurchatov Institute”, 123182, Moscow, Russia}\\
 $^d$\textit{Department of Particle Physics and Cosmology, Physics Faculty,\\
  M.V. Lomonosov Moscow State University,\\
  Vorobjevy Gory, 119991 Moscow, Russia}
\end{center}

\vspace{5pt}

\begin{abstract}
Considering an anisotropic cosmological background is an interesting and simultaneously challenging problem of theoretical physics, since we not only assume a high degree of anisotropy in the early stages of the Universe, but also observe it to a small degree until now. In this paper we have constructed the unconstrained action for the perturbations above Bianchi I type background in the most general scalar-tensor theory of gravity, the Horndeski theory, and evaluate the effect of the deviation from the anisotropic background  on the previously established stable solution obtained in previous works.
\end{abstract}
\section{Introduction}

Horndeski theory~\cite{Horndeski:1974wa,Nicolis:2008in,Deffayet:2009wt,Fairlie:1991qe}(See \cite{Kobayashi:2019hrl} for a review) is the most general scalar-tensor theory of gravity with second derivatives in the equations of motion, which makes it interesting in matters of constructing new models of dark matter, dark energy, wormholes, and so on. However, the construction of stable solutions to the theory is limited by the so-called no-go theorem~\cite{Libanov:2016kfc,Kobayashi:2016xpl}. It was formulated in the case of a homogeneous and isotropic cosmological background and for a spherically symmetric background. The options for avoiding it were also widely considered~\cite{Cai:2016thi,Creminelli:2016zwa,Kobayashi:2016xpl,Kolevatov:2017voe,Cai:2017dyi,Cai:2017tku,Mironov:2018oec,Mironov:2019qjt,Mironov:2019mye,Ilyas:2020qja, Ilyas:2020zcb, Ageeva:2020buc,Ageeva:2020gti,Ageeva:2021yik,Mironov:2022quk}.

However, the list of problems considered in the context of Horndeski theory does not end here, and there are papers that consider anisotropic background~\cite{Starobinsky:2019xdp,Galeev:2021xit}. From a cosmological point of view, it is generally assumed that the universe was highly anisotropic in its early inflation stages~\cite{Belinsky:1970ew,Collins:1972tf,Belinsky:1982pk} and in addition, some degree of anisotropy is also present in current observations. In this context, studying and constructing potentially non-singular solutions in the case of an anisotropic background becomes relevant. 

In the previously considered models there exists a no-go theorem, which prohibits stable solutions in the general Horndeski theory on the whole time axis. However, the theorem is formulated for isotropic background, so the question arises whether the existence of this theorem is a consequence of high symmetry of spacetime. This question requires further study. 
On the other hand, the authors of the paper considered a way to circumvent the no-go theorem in the isotropic case~\cite{Mironov:2022quk}. In the fact that we consider lagrangian with special relation between function, so, that only scalar degree of freedom is non-dynamical above homogeneous isotropic background and no-go theorem is not applicapable. Earlier for this situation stable solutions of Bounce and Genesis types were constructed. In this paper we show that this way no longer works, and the solution is actually unstable. This becomes apparent if we introduce a small anisotropy into the existing solution. The result is actually important, as it illustrates that trying to get rid of one of the degrees of freedom of a system over a particular background sometimes leads to pathologies over nearby solutions.

In this paper, we consider perturbations of the metric and scalar field above a homogeneous but anisotropic Bianchi type I background. We integrate out the non-dynamical variables in partly gauge invariant form and construct the unconstrained action for the perturbations. We also show how the quadratic action reduces to the isotropic case of the Friedmann universe and checked how the deviation from the isotropic background affects the previously obtained\cite{Mironov:2022quk} stable solution.

\section{Perturbations about anisotropic background}\label{perturbations}

We consider the Horndeski theory with the following Lagrangian:
\begin{subequations}
 \label{lagrangian}
 \[S=\int\mathrm{d}^4x\sqrt{-g}\left(\mathcal{L}_2 + \mathcal{L}_3 + \mathcal{L}_4\right),\]
 \vspace{-0.8cm}
 \begin{align}
   & \mathcal{L}_2=F(\pi,X),                                                                                  \\
   & \mathcal{L}_3=K(\pi,X)\Box\pi,                                                                           \\
   & \mathcal{L}_4=-G_4(\pi,X)R+2G_{4X}(\pi,X)\left[\left(\Box\pi\right)^2-\pi_{;\mu\nu}\pi^{;\mu\nu}\right],
 \end{align}
\end{subequations}
where $\pi$ is the scalar field, $X=g^{\mu\nu}\pi_{,\mu}\pi_{,\nu}$, $\pi_{,\mu}=\partial_\mu\pi$, $\pi_{;\mu\nu}=\triangledown_\nu\triangledown_\mu\pi$, $\Box\pi = g^{\mu\nu}\triangledown_\nu\triangledown_\mu\pi$, $G_{4X}=\partial G_4/\partial X$, etc.

In this paper we consider anisotropic background:
\[ ds^{2} = dt^2 - \left(a^2(t) dx^2 + b^2(t) dy^2 + c^2(t)dz^2\right). \]
The decomposition of metric perturbations $h_{\mu \nu}$ into helicity components in this case has the form
\begin{subequations}
 \begin{align}
   & h_{00}=2 \Phi \\
   & h_{0 i}= - \partial_{i} \beta + Z_i^T,                                                                                           \\
   & h_{i j}=-2 \frac{H_i}{H} \Psi g_{i j}-2 \partial_{i}\partial_{j} E - \left(\partial_{i} W_j^T+\partial_{j} W_i^T\right)+h_{i j}^{TT},
 \end{align}
\end{subequations}
where $\Phi, \beta, \Psi, E$ are scalar fields, $H_i$ is the corresponding Hubble parameter (here and further $i = a, b, c$) and $H = \dfrac{1}{3} \left( H_a + H_b + H_c\right) $, $Z_{i}^{T}, W_{i}^{T}$ are transverse vector fields ($\partial_{i}Z_{i}^{T} = \partial_{i} W_{i}^{T} = 0$), $h^{TT}_{ij}$ is transverse traceless tensor. We also denote the perturbation of scalar field $\delta \pi = \chi$.

The action for tensor perturbations has the form:
\[\label{tensor_action} S_h^{(2)}=\int \mathrm{d} t \mathrm{~d}^3 x a^3\left[\frac{A_5}{2} \left(\dot{h}_{i j}\right) ^2-{A_2}\left( \Delta_a^2 h^{TT}_{i j}+ \Delta_b^2 h^{TT}_{i j} + \Delta_c^2 h^{TT}_{i j}\right)\right].\]

Here dot denotes the derivative with respect to the cosmic time $t$, $\Delta_a = a^{-1} \partial_{x}, \Delta_b = b^{-1} \partial_{y}, \Delta_c = c^{-1} \partial_{z}$,  coefficients $A_i$ are the combinations of the Lagrangian functions and their derivatives.

Similar to the isotropic case, vector perturbations are non-dynamical and scalars are the most challenging. Without loss of generality we partly use the gauge freedom and gauge away the longitudinal component $\partial_{i}\partial_{j} E$ from the very beginning. Then the second-order action for scalar sector has the form:

\begin{align}\label{action}
  &S^{(2)} = \int {\rm d} x \,\, a b c \left( \frac{1}{6} {A_{1}}\sum_{i\neq j}  \dot{\Psi}_i \dot{\Psi}_j  + \frac{A_{2}}{2} \sum_{\substack{i=a,b,c\\ i\neq j\neq k}} \Delta_i \Psi_j \Delta_i \Psi_k +{A_{3}} {\Phi}^{2} \right. & \nonumber\\[2ex]
  & +\Phi \left(A_{4}^{i} \Delta_i^2 \beta\right) +A_5 \sum_{\substack{i=a,b,c\\ i\neq j\neq k}} \dot{\Psi}_i \left(\Delta_j^2 \beta + \Delta_k^2 \beta \right) +\Phi \left(A_{6}^{i}\dot{\Psi}_i\right) +\frac{A_7}{2} \Phi \sum_{\substack{i=a,b,c\\ i\neq j\neq k}} \Delta_i^2 \left( \Psi_j + \Psi_k \right)& \nonumber\\[2ex]
  &+\Phi \left(A_{8}^{i} \Delta_i^2 \chi\right) +\dot{\chi} \left(A_{9}^{i} \Delta_i^2 \beta\right) + \chi \left( A_{10}^{i} \ddot{\Psi}_i\right) +{A_{11}} \Phi \dot{\chi}+ \chi \left(A_{12}^{i} \Delta_i^2 \beta\right) &\nonumber\\[2ex]
  &  + \chi \sum_{i,j} \frac12 A_{13}^{ij} \left(\Delta_i^2 \Psi_j + \Delta_j^2 \Psi_i\right)+{A_{14}} {\left(\dot{\chi}\right)}^{2} +A_{15}^{i} \left(\Delta_i \chi\right)^2  + {A_{17}} \Phi \chi+ \chi \left(A_{18}^{i} \dot{\Psi}_i\right) &\nonumber\\[2ex]
  &    +{A_{20}} {\chi}^{2} + \frac12 \sum_{\substack{i,j=a,b,c\\ i\neq j}} B^{ij} \Psi_i \dot{\Psi}_j - \Psi_a \left(B^{ab} \Delta_y^2 \beta + B^{ac} \Delta_z^2 \beta \right)+\Psi_b \left(B^{ab} \Delta_x^2 \beta + B^{bc} \Delta_z^2 \beta \right) &\nonumber\\[2ex]
  & \left. +\Psi_c \left(B^{ac} \Delta_x^2 \beta-B^{bc} \Delta_y^2 \beta \right) \right).&
\end{align}
Here $\Psi_i = \bar{H}_i \Psi$ and $\bar{H}_i = H_i/H$ and we assume summation by dummy indices. The explicit form of these coefficients can be found in the Appendix A\footnote{Note that all coefficients in before the spatial derivatives, except for $A_1$ and $A_2$, suffer a "splitting". This is due to the anisotropy of the background in consideration. The coefficients $A_1$ and $A_2$ also undergoe "splitting", but only when $\mathcal{L}_5$ summand of the general Horndeski theory is considered.}. The choice of notations $A_i$ is related to the attempt to show the correspondence between the isotropic and anisotropic cases; the coefficients $B^{ij}$, in turn, correspond to the terms that do not exist in the isotropic case. The use of auxiliary variables $\Psi_i$ is due to the convenience and shortness of the presentation.

\section{Gauge invariant variables}

In this section, we reduce \eqref{action} to a single variable using the gauge invariant variables.

Because we have already partially fixed the gauge by putting $\partial_{i}\partial_{j} E = 0$, action~\eqref{action} is invariant under the residual gauge transformations:
\[\label{gauge} \Phi \to \Phi + \dot{\xi}_0,\quad \beta \to \beta - \xi_0, \quad \chi \to \chi + \xi_0\dot{\pi},\quad \Psi \to \Psi + \xi_0 H.\]

Based on this, we can introduce gauge-invariant variables:
\begin{subequations}
  \begin{align}
    &\X  = \chi + \dot{\pi} \beta, &\\
    &\Y_i = \Psi_i + H_i \beta,&\\
    &\Z = \Phi + \dot{\beta}.&
  \end{align}
\end{subequations}

In terms of these variables, the action is rewritten as follows:
\begin{align}
  &S^{(2)} =\int{\rm d}x \,\, a b c \left(\frac{1}{6}{A_{1}} \sum_{i\neq j} \dot{\Y}_i \dot{\Y}_j +\frac{A_{2}}{2} \sum_{\substack{i=a,b,c\\ i\neq j\neq k}} \Delta_i \Y_j \Delta_i \Y_k  + {A_{3}} {\Z}^{2} + \Z \left(A_{6}^{i} \dot{\Y}_i\right) \right.& \nonumber \\
 &   + \frac{A_7}{2} \Z \sum_{\substack{i=a,b,c\\ i\neq j\neq k}} \Delta_i^2 \left( \Y_j + \Y_k \right) + \Z \left(A_{8}^{i} \Delta_i^2 \X \right) + \X \left( A_{10}^{i}\ddot{\Y}_i\right) +{A_{11}} \Z \dot{\X} &\nonumber\\
  &  + \X \sum_{i,j} \frac12 A_{13}^{ij} \left(\Delta_i^2 \Y_j + \Delta_j^2 \Y_i\right) +{A_{14}} \left({\dot{\X}}\right)^{2} +\left(A_{15}^{i} \left(\Delta_i \X \right)^2 \right)+{A_{17}} \X \Z &\nonumber\\
  &\left. + \X \left(A_{18}^{i}\dot{\Y}_i\right) +{A_{20}} {\X}^{2} + C_{3}^{ab} \Y_a \dot{\Y}_b+C_{3}^{bc} \Y_c \dot{\Y}_b+C_{3}^{ac} \Y_a \dot{\Y}_c \right). &
\end{align}

The variable $\Z$ is obviously non-dynamical and we can derive a $\Z$–constraint which has the following form:
\begin{align}
  &\Z = - \frac12 \frac{1}{A_3} \left(\left(A_{8}^{i} \Delta_i^2 \X\right) + A_6^i \dot{\Y}_i   + {A_{11}} \dot{\X} \frac{A_7}{2} \sum_{\substack{i=a,b,c\\ i\neq j\neq k}} \Delta_i^2 \left( \Y_j + \Y_k \right) + \X {A_{17}}\right).&
\end{align}

Then we substitute $\Y_i = \bar{H_i} \Y$ and replace the variables as linear combination:
\[\label{zeta} \zeta = \Y + \eta \X,\text{ where }\eta = 9\dfrac{{A_{11}} {A_{4}}+2{A_{3}} {A_{8}}}{\left(4\left(\bar{H}_a \bar{H}_b+\bar{H}_a \bar{H}_c+\bar{H}_b \bar{H}_c\right) {A_{1}} {A_{3}}-27{{A_{4}}}^{2}\right)} ,\]
here and further
\[A_4 = \frac13 \sum_{l = a,b,c} A_4^l \bar{H}_l, \quad A_8 = \frac13 \sum_{l = a,b,c} A_8^l \bar{H}_l.\]

 The introduction of the new variable $\zeta$ is intended to distinguish the explicitly non-dynamic variable $\X$. This is not the only way to introduce the dynamical variable, we can also leave $\Y$ to be a constraint. Physically, of course, the result will not depend on the choice of the variables, but the form of the action may change.

The action in terms of the variables $\X$ and $\zeta$ is as follows:
\begin{align}\label{2varAction}
  &S^{(2)} = \int {\rm d}x \,\, abc \left({\left(\dot{\zeta}\right)}^{2} \left( \frac23 {A_{1}} \left(\bar{H}_a \bar{H}_b+\bar{H}_a \bar{H}_c + \bar{H}_b \bar{H}_c\right) - \frac{9}{2} \frac{{{A_{4}}}^{2}}{{A_{3}}} \right)-\dot{\zeta} \mathcal{X} {C_{3}} \right.&\nonumber\\
  &\left. - \frac{1}{2 A_3} {\left(\mathcal{X} {C_{1}}-\zeta {C_{2}}\right)}^{2}+M {\mathcal{X}}^{2}+{\zeta}^{2} \left(m+{C_{4}}\right)+\mathcal{X} \zeta {C_{5}}\right),&
\end{align}
where
\begin{subequations}
  \begin{align}
    &{C_{1}} = \Sigma_a \frac{k_x^2}{a^2}+\Sigma_b \frac{k_y^2}{b^2}+\Sigma_c \frac{k_z^2}{c^2},&\\
    &{C_{2}} = \Theta_a \frac{k_x^2}{a^2}+\Theta_b \frac{k_y^2}{b^2}+\Theta_c \frac{k_z^2}{c^2},&
    \end{align}
    \begin{align}
    &{C_{3}} = \Lambda_a \frac{k_x^2}{a^2} +\Lambda_b \frac{k_y^2}{b^2}+\Lambda_c \frac{k_z^2}{c^2},&\\
    &{C_{4}} = \Pi_a \frac{k_x^2}{a^2} +\Pi_b \frac{k_y^2}{b^2}+ \Pi_c \frac{k_z^2}{c^2},&\\
    &{C_{5}} = \Xi_a \frac{k_x^2}{a^2} +\Xi_b \frac{k_y^2}{b^2}+ \Xi_c \frac{k_z^2}{c^2},&
  \end{align}
\end{subequations}
where $\Lambda_i, \Xi_i, \Pi_i, \Theta, M, m$ and $\Sigma_i$ are the linear combinations of $A_i$, which can be found in Appendix B.

As we said above, it is now clearly seen that the variable $\X$ is non-dynamical and the corresponding constraint is:
\[\label{Xconstr}\X = \frac{1}{{{{C_{1}}}^{2}-2M {A_{3}}}} \left(\left({C_{1}} {C_{2}}+{A_{3}} {C_{5}}\right) \zeta  -{A_{3}} {C_{3}} \dot{\zeta}\right) \]

Upon substitution \eqref{Xconstr} into \eqref{2varAction}, we get
\begin{align}\label{finishAction}
  &S^{(2)} = \int {\rm d}x \,\, a b c \left({\left(\dot{\zeta}\right)}^{2} \left(\frac{2}{3}{A_{1}}\left( \bar{H}_a \bar{H}_b+\bar{H}_a \bar{H}_c+\bar{H}_b \bar{H}_c\right) - \frac{9}{2} \frac{{{A_{4}}}^{2}}{A_3} + \frac{1}{2} \frac{{A_{3}} {{C_{3}}}^{2}}{{{{C_{1}}}^{2}-2M {A_{3}}}} \right)+\right.&\nonumber\\[2ex]
  &  + {\zeta}^{2} \left({C_{4}} +\frac{{C_{1}} {C_{2}} {C_{5}}}{{{{C_{1}}}^{2}-2M {A_{3}}}} +\frac{1}{2} \frac{1}{abc}\left[a b c \frac{{C_{1}} {C_{2}} {C_{3}}}{{{{C_{1}}}^{2}-2M {A_{3}}}}\right] \right)+&\nonumber\\[2ex]
  & \left.+{\zeta}^{2}  \left(m+\frac{1}{2}\frac{{A_{3}}{{C_{5}}}^{2} + 2 {{C_{2}}}^{2} M}{{{{C_{1}}}^{2}-2M {A_{3}}}} +\frac{1}{2}\frac{1}{abc}\frac{d}{dt}\left[\frac{a b c {A_{3}} {C_{3}} {C_{5}}}{{{{C_{1}}}^{2}-2M {A_{3}}}}\right]\right) \right) .&
\end{align}

In action~\eqref{finishAction} the first term is the kinetic term, the second one is gradient term (proportional to $k^2$), and the third one corresponds to the mass of the field $\zeta$.

\section{Isotropic limit}

In the isotropic case $b = a, c = a$ the coefficients $C_i$, respectively, change to:
\begin{subequations}
  \begin{align}
    &C_1 = A_4 \frac{\left(2{A_{1}} {A_{11}}+9{A_{4}}{A_{8}}\right)} {\left(4{A_{1}} {A_{3}}-9{{A_{4}}}^{2}\right)} \frac{k^2}{a^2},&\\
    &C_2 = \frac23 A_1 \frac{k^2}{a^2},&\\
    &C_3 = \frac{9{A_{4}} {A_{8}}+2{A_{1}} {A_{11}}}{3 {A_{3}}} \frac{k^2}{a^2},&\\
    &C_4 = \left(A_2 +  \frac12 \frac{1}{a} \frac{d}{dt} \left[\frac{A_4 A_1 a}{A_3}\right]\right) \frac{k^2}{a^2},&\\
    &C_5 = 0,&\\
    &m = M = 0.&
  \end{align}
\end{subequations}

And action~\eqref{finishAction} takes the form:
\begin{equation}\label{unitary_action}
    S^{(2)} = \int \mathrm{d}t\,\mathrm{d}^3x\,a^3  \left({\mathcal{G}_S \left(\dot{\zeta}\right)}^{2} - \mathcal{F}_S \dfrac{\left(\overrightarrow{\nabla} \zeta\right)^2}{a^2} \right),
\end{equation}
where
\begin{subequations}
    \begin{align}
        \mathcal{G}_S &= \frac{4}{9}\frac{{A_{3}} {{A_{1}}}^{2}}{{A_{4}}^2}-{A_{1}},\\
        \mathcal{F}_S &= -\frac1a \frac{\mathrm{d}}{\mathrm{d}t}\left[\frac{a A_{1} A_{7}}{3 A_{4}}\right]- {A_{2}} = \dfrac{1}{a} \frac{\mathrm{d}}{\mathrm{d}t}\left[ \frac{a A_5 \cdot A_7}{2 A_4}\right] - A_2.
    \end{align}
\end{subequations}
which corresponds to the known result.

\section{Testing the stability of the Bounce solution with respect to small anisotropy} 

To further analyze the theory, we consider the action~\eqref{action} in the unitary gauge $\chi=0$ and direct the momentum $\vec{k}$ along the x-axis, so $\vec{k} = (k_x, 0, 0)$. Then by varying the action with respect to the fields $\Phi$ and $\beta$ we obtain the following constraint equations

\begin{subequations}
  \begin{align}
    \Phi &= \frac{A_1}{3 A_{4}^{x}} \left(\dot{\Psi_b}+\dot{\Psi_c} - \left(H_a-H_b\right) \Psi_b - \left(H_a-H_c\right) \Psi_c\right)\\
    k_x^2 \beta &= \frac{1}{A_4^x} \left( \left(\dot{\Psi}_i A_4^i\right) - \frac13 A_1 k_x^2 \left(\Psi_b + \Psi_c\right)\right) \nonumber \\
    &+ \frac{2}{3} \frac{A_1 A_3}{\left(A_4^x\right)^2} \left(\left(H_a - H_b\right) \Psi_b + \left(H_a - H_c\right) \Psi_c - \left(\dot{\Psi}_b + \dot{\Psi}_c\right)\right)
  \end{align}
\end{subequations}
After removing the constraints, we get the following action on the $\Psi$ variable
\begin{equation}\label{unitary_anisotropic_action}
    S^{(2)} = \int \mathrm{d}t\,\mathrm{d}^3x\, a b c  \left({\mathcal{G}_S \left(\dot{\Psi}\right)}^{2} + M \Psi^2 + \mathcal{F}_S \frac{k_x^2}{a^2}  \Psi^2 \right),
\end{equation}
where
\begin{subequations}
  \begin{align}
    &\mathcal{G}_{S} = \frac{2}{9}\frac{{A_{3}} {{A_{1}}}^{2}}{\left(A_{4}^{x}\right)^2} {\left(\bar{H}_b+\bar{H}_c\right)}^{2}-\frac23 \frac{{A_{1}} }{{A_{4}^{x}}} \left(A_{4}^{y} \bar{H}_b+A_{4}^{z} \bar{H}_c\right) \left(\bar{H}_b+\bar{H}_c\right)+\frac{2}{3}{A_{1}} \bar{H}_b \bar{H}_c,&\\
    &\mathcal{F}_{S} = -2A_{2} \bar{H}_b \bar{H}_c - \frac{1}{9 a^3 } {\left(\bar{H}_b+\bar{H}_c\right)}^{2} \frac{d}{dt} \left[\frac{A_1^2 a^3}{A_4^x}\right] + \frac{{{A_{1}}}^{2}}{9 {A_{4}^{x}}} \left({\bar{H}_b}^{2}-{\bar{H}_c}^{2}\right) \left(H_b-H_c\right),&
  \end{align}
\end{subequations}
the explicit value of $M$ is not important to us now. Even though we resolved the constraints again in unitary gauge, the speed of sound coincides with one which we obtained from the action \eqref{finishAction}.

\begin{figure}[H]\begin{center}\hspace{-1cm}
{\includegraphics[width=1\linewidth]{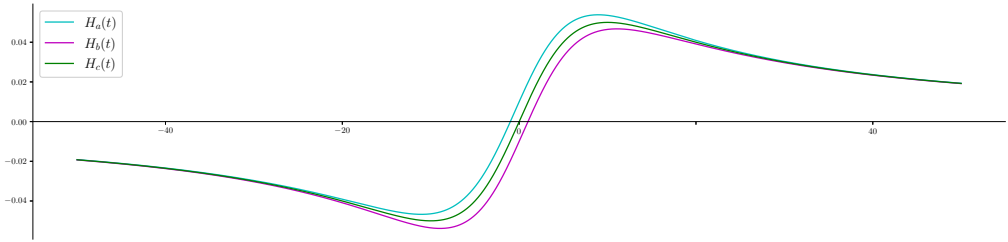}}
\caption{\footnotesize{Hubble parametrs $H_a(t), H_b(t), H_c(t)$, when we choose $\alpha=10, \tau = 10$. The bounce solution is characterized by the presence of a twist at the origin and tending to 0 at $\pm$ infinity.}} 
\label{hubbles}
\end{center}\end{figure}

Let us now consider a model with the following Lagrangian
\[\label{A4=0_Lagrangian} \mathcal{L}=\frac{\pi^2-\tau^2}{3\left(\tau^2+\pi^2\right)^2}-\frac{\pi^2 X}{\left(\tau^2+\pi^2\right)^2}+\frac{\pi X}{3\left(\tau^2+\pi^2\right)} \square \pi+\frac{1}{2} R.\]
This Lagrangian was obtained in the Ref.\cite{Mironov:2022quk} and corresponds to the situation where in the action \eqref{unitary_action} $A_4 = 0$. This model has an isotropic bounce solution with out dynamical scalar modes. Let us check whether the solution remains stable if we consider an anisotropic bounce - deviate from the isotropic case in two directions:
\[H_a = \frac{t}{\left({\tau}^{2}+{t}^{2}\right)} + \frac{\alpha}{\left({\tau}^{2}+{t}^{2}\right)^{3/2}}, \quad H_b = \frac{t}{\left({\tau}^{2}+{t}^{2}\right)} - \frac{\alpha}{\left({\tau}^{2}+{t}^{2}\right)^{3/2}}, \quad H_c = \frac{t}{\left({\tau}^{2}+{t}^{2}\right)}. \]
Here the parameter $\tau$ defines the bounce amplitude and $\alpha$ the degree of deviation from the isotropic case (See Pic.\ref{hubbles}). 

To analyze the stability of the scalar field, we numerically plot the square of the speed of sound $c_S^2$(\ref{alpha1_tau20_t10}):

\begin{figure}[H]\begin{center}\hspace{-1cm}
{\includegraphics[width=0.5\linewidth]{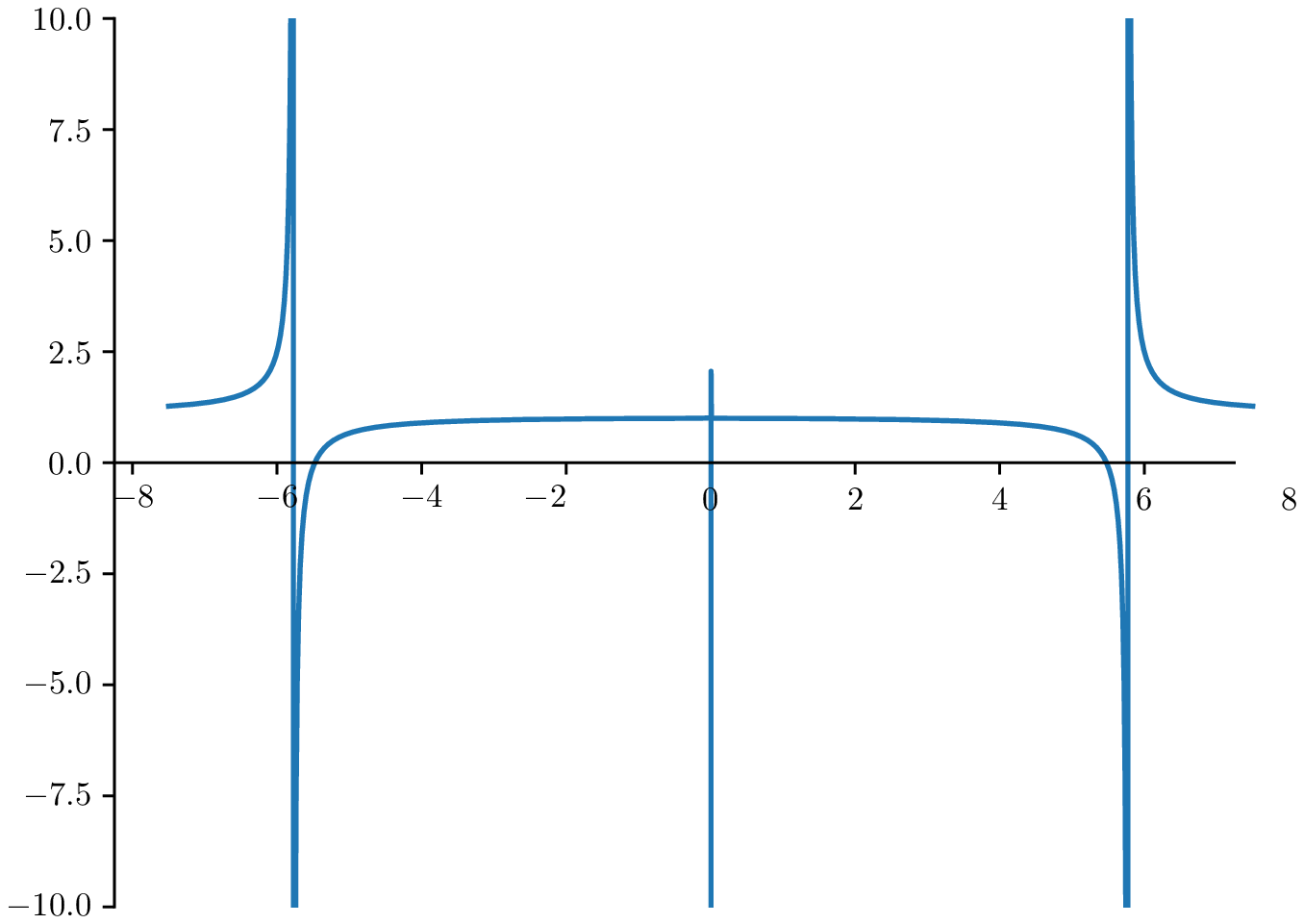}}\hspace{2.8cm}\hspace{-3cm}
{\includegraphics[width=0.5\linewidth]{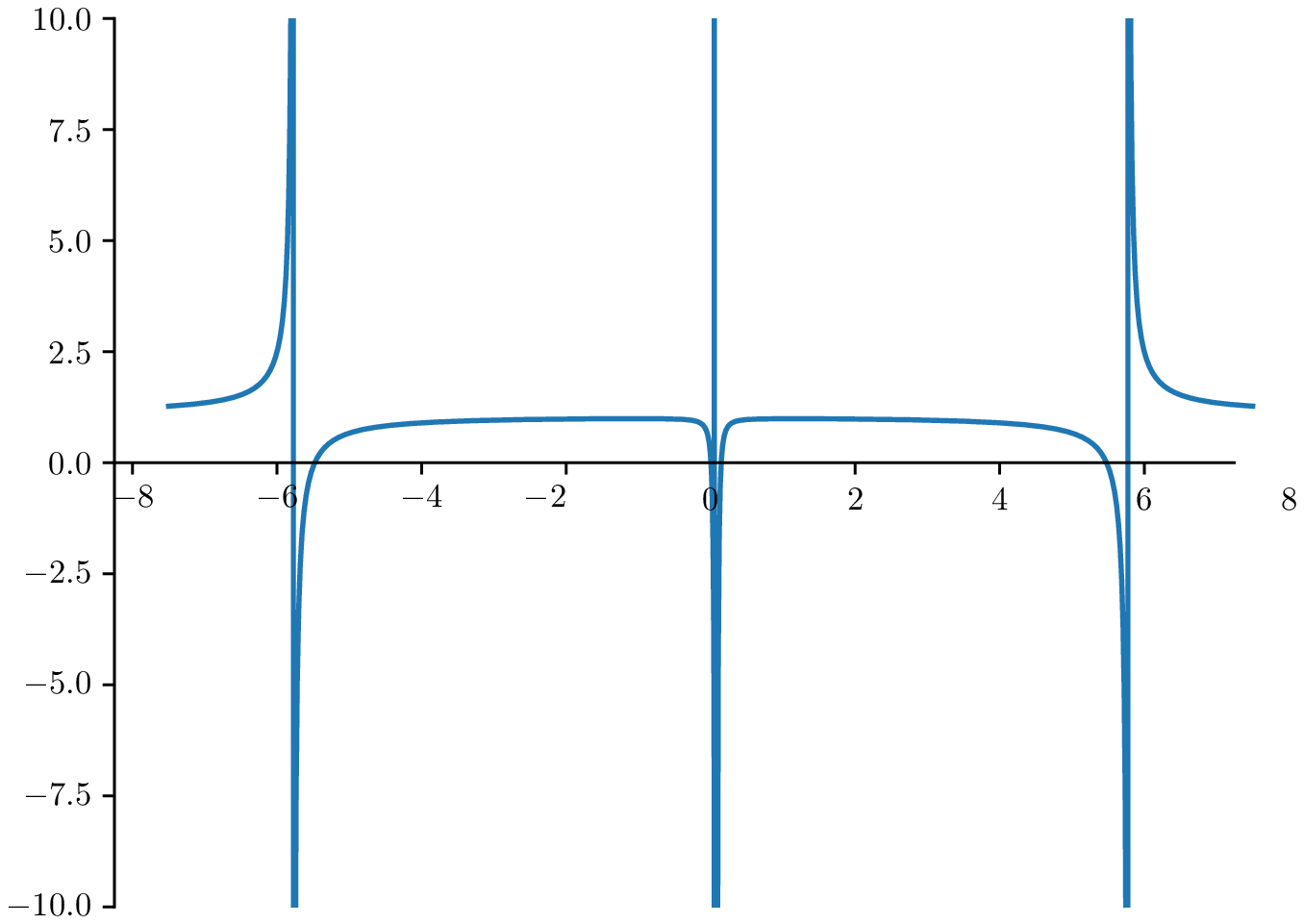}}
\caption{\footnotesize{The square of the speed of sound $c_S^2$, when we choose $\alpha=0.1, \tau = 10$ (left panel) and $\alpha = 1, \tau = 20$ (right panel). In this case, the square of the speed of sound will have at least 2 symmetric singular points and tends to 0 as univerce becomes isotropic.}} 
\label{alpha1_tau20_t10}
\end{center}\end{figure}

\begin{figure}[H]\begin{center}\hspace{-1cm}
{\includegraphics[width=0.5\linewidth]{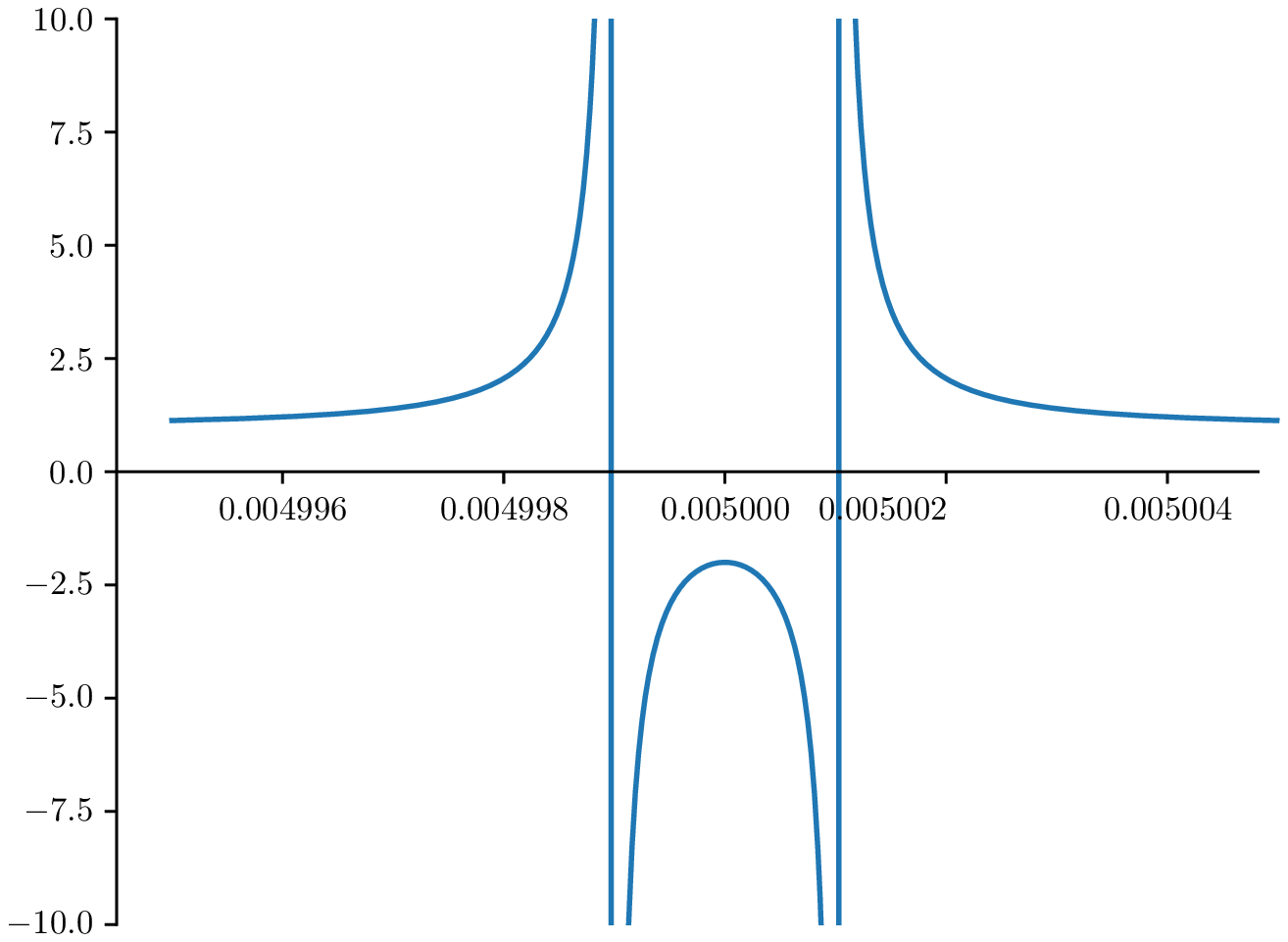}}\hspace{2.8cm}\hspace{-3cm}
{\includegraphics[width=0.5\linewidth]{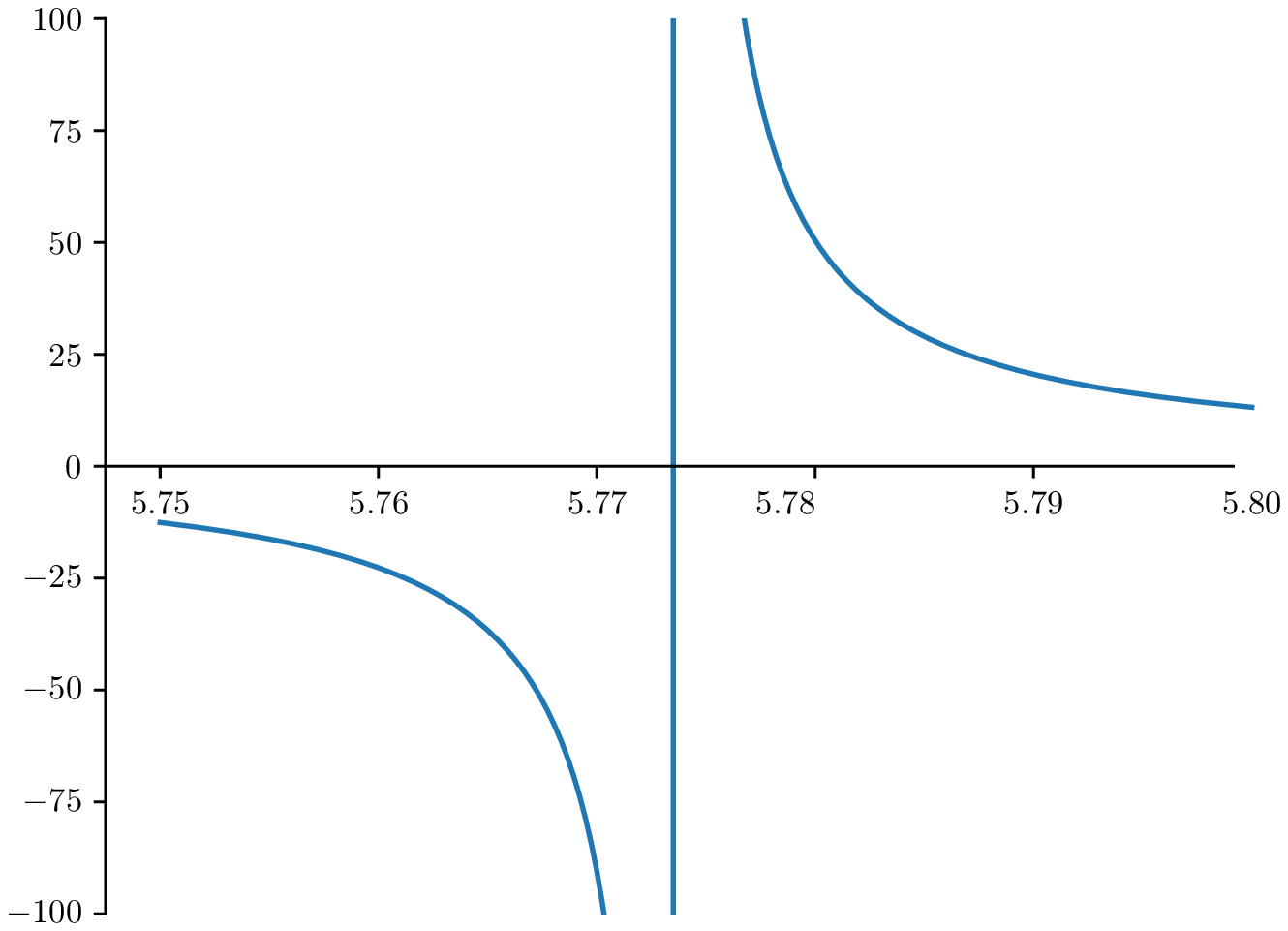}}
\caption{\footnotesize{Zoom of the neighborhood of singular points of the square of the speed of sound $c_S^2$ for parameters $\alpha=0.1, \tau = 10$.}} 
\label{alpha1_tau20_t10zoom}
\end{center}\end{figure}

The plots Pic.\ref{alpha1_tau20_t10zoom} show that in the theory \eqref{A4=0_Lagrangian} the scalar field becomes unstable even with a small deviation from the isotropic background. This tells us that the result obtained earlier in Ref.\cite{Mironov:2022quk} is a very special case directly related to the background isotropy. 
\section{Conclusion}

In this paper, we construct an action for tensor and scalar modes of perturbations of the metric and scalar fields over an anisotropic background, and check whether the previously obtained solution for the universe with bounce remains stable. It turned out that the stability of perturbations in the universe with bounce is directly related to its isotropy and even at small deviations from it the square of the speed of sound diverges and become negative. However, our result still opens a wide avenue for finding stable solutions and supports the quest for power spectra of anisotropic models of the early Universe.
\section*{Acknowledgements}
{The authors wish to thank Kasper Peeters for developing and maintaining cadabra2 software~\cite{Peeters:2018dyg}, with which most of the calculations were performed.} This work has been supported by Russian Science Foundation grant 19-12-00393.

\newpage
\input{Supplemental_Material/Supplemental}
\newpage
\printbibliography
\end{document}

%% file: Supplemental_Material/Supplemental.tex
\section*{Appendix A}

In this Appendix we collect the expressions for coefficients $A_i$
entering the quadratic action. Here and below the small Latin indices run the values $a,b,c$ and we consider that $i \neq j \neq k$:
\begin{flalign}
  & {A_{1}} = -6{G_{4}}+12G_{4X} {\left(\left(\dot{\pi}\right)\right)}^{2}, &
\end{flalign}
\begin{flalign}
  & A_{2} = 2{G_{4}}, &
\end{flalign}
\begin{flalign}
  & {A_{3}} = F_{X} {\left(\left(\dot{\pi}\right)\right)}^{2}+2F_{X X} {\left(\left(\dot{\pi}\right)\right)}^{4}+4\left(H_a+H_b+H_c\right) K_{X} {\left(\left(\dot{\pi}\right)\right)}^{3} -K_{\pi} {\left(\left(\dot{\pi}\right)\right)}^{2} & \nonumber\\
  & -K_{X \pi} {\left(\left(\dot{\pi}\right)\right)}^{4}+2\left(H_a+H_b+H_c\right) K_{X X} {\left(\left(\dot{\pi}\right)\right)}^{5}-2 \left(H_a H_b + H_a H_c + H_b H_c\right) {G_{4}}                                                       & \nonumber\\
  & +14\left(H_a H_b+H_a H_c+H_b H_c\right) G_{4X} {\left(\left(\dot{\pi}\right)\right)}^{2}+32 \left(H_a H_b+H_a H_c+H_b H_c\right) G_{4XX} {\left(\left(\dot{\pi}\right)\right)}^{4}                                                        & \\
  & -10\left(H_a+H_b +H_c\right) G_{4X \pi} {\left(\left(\dot{\pi}\right)\right)}^{3} -2 \left(H_a+H_b+H_c\right) G_{4\pi} \left(\dot{\pi}\right)                                                                                             & \nonumber\\
  & +8\left(H_a H_b+H_a H_c+H_b H_c\right) G_{4XXX} {\left(\left(\dot{\pi}\right)\right)}^{6}-4\left(H_a+H_b+H_c\right) G_{4 X X \pi } {\left(\left(\dot{\pi}\right)\right)}^{5},\nonumber                                                             &
\end{flalign}
\begin{flalign}
 &A_{4}^{i} = 2K_{X} {\left(\left(\dot{\pi}\right)\right)}^{3}-2\left(H_j+H_k\right) {G_{4}}+8\left(H_j+H_k\right) G_{4X} {\left(\left(\dot{\pi}\right)\right)}^{2}&\nonumber\\
 &-4G_{4X \pi} {\left(\left(\dot{\pi}\right)\right)}^{3}-2G_{4\pi} \left(\dot{\pi}\right)+8\left(H_j+H_k\right) G_{4XX} {\left(\left(\dot{\pi}\right)\right)}^{4},
\end{flalign}
\begin{flalign}
  &A_5 = -\frac{1}{3} A_1&\\
  &A_6^i = - A_4^i&\\
  &A_7 = \frac13 A_1&
\end{flalign}
\begin{flalign}
  &A_{8}^{i} = 2K_{X} {\left(\left(\dot{\pi}\right)\right)}^{2}-2G_{4\pi}+4\left(H_j+H_k\right) G_{4X} \left(\dot{\pi}\right)+8\left(H_j+H_k\right) G_{4XX} {\left(\left(\dot{\pi}\right)\right)}^{3}&\\
  &-4G_{4X \pi} {\left(\left(\dot{\pi}\right)\right)}^{2},&\nonumber
\end{flalign}
\begin{flalign}
  &A_9^i = - A_8^i &\\
  & A_{10}^i = - A_8^i&
\end{flalign}
\begin{flalign}
  & {A_{11}} = -2F_{X} \left(\dot{\pi}\right)-4F_{X X} {\left(\left(\dot{\pi}\right)\right)}^{3}+2K_{X \pi} {\left(\left(\dot{\pi}\right)\right)}^{3}+2K_{\pi} \left(\dot{\pi}\right)-4 \left(H_a+H_b+H_c\right) K_{X X} {\left(\left(\dot{\pi}\right)\right)}^{4}\nonumber & \\
  & -6\left(H_a+H_b+H_c\right) K_{X} {\left(\left(\dot{\pi}\right)\right)}^{2}-12\left(H_b H_c+H_a H_c+H_a H_b\right) G_{4X} \left(\dot{\pi}\right)                                                                                                                \nonumber& \\
  & -48\left(H_b H_c+H_a H_c+H_a H_b\right) G_{4XX} {\left(\left(\dot{\pi}\right)\right)}^{3}+16\left(H_a+H_b+H_c\right) G_{4X \pi} {\left(\left(\dot{\pi}\right)\right)}^{2}                                                                                      & \\
  & +2\left(H_a+H_b+H_c\right) G_{4\pi}-16\left(H_a H_b+H_a H_c+H_b H_c\right) G_{4XXX} {\left(\left(\dot{\pi}\right)\right)}^{5}                                                                                                                                 \nonumber & \\
  & +8\left(H_a+H_b+H_c\right) G_{4 X X \pi } {\left(\left(\dot{\pi}\right)\right)}^{4},\nonumber                                                                                                                                                                           &
\end{flalign}
\begin{flalign}
  & A_{12} = 2F_{X} \left(\dot{\pi}\right)+2\left(H_a+H_b+H_c\right) K_{X} {\left(\left(\dot{\pi}\right)\right)}^{2}-2K_{\pi} \left(\dot{\pi}\right) & \nonumber\\
  & +4\left(H_b H_c+H_a H_c+H_a H_b\right) G_{4X} \left(\dot{\pi}\right)+2G_{4\pi\pi} \left(\dot{\pi}\right)-2G_{4\pi} H_a                           & \\
 &+8\left(H_a H_b+H_a H_c+H_b H_c\right) G_{4XX} {\left(\left(\dot{\pi}\right)\right)}^{3} -8 \left(H_a + H_b+ H_c\right) G_{4X \pi} {\left(\left(\dot{\pi}\right)\right)}^{2},\nonumber &                                                                                                                  \end{flalign}
\begin{flalign}
  &A_{13}^{ij} = -2G_{4\pi}+4G_{4X} \left(\ddot{\pi}\right)+4G_{4X} H_k \left(\dot{\pi}\right)+4G_{4X \pi} {\left(\left(\dot{\pi}\right)\right)}^{2}+8G_{4XX} \left(\ddot{\pi}\right) {\left(\left(\dot{\pi}\right)\right)}^{2}&
\end{flalign}
\begin{flalign}
  & {A_{14}} = 2F_{X X} {\left(\left(\dot{\pi}\right)\right)}^{2}+F_{X}-K_{\pi}-K_{X \pi} {\left(\left(\dot{\pi}\right)\right)}^{2}+2\left(H_a+H_b+H_c\right) K_{X X} {\left(\left(\dot{\pi}\right)\right)}^{3} \nonumber& \\
  & +2\left(H_a+H_b+H_c\right) K_{X} \left(\dot{\pi}\right)-6\left(H_a+H_b+H_c\right) G_{4X \pi} \left(\dot{\pi}\right)                                                                                         & \\
  & +16\left(H_a H_b+H_a H_c+H_b H_c\right) G_{4XX} {\left(\left(\dot{\pi}\right)\right)}^{2}+2\left(H_a H_b+H_a H_c+H_b H_c\right) G_{4X},                                                                 \nonumber    & \\
  & -4\left(H_a+H_b+H_c\right) G_{4 X X \pi } {\left(\left(\dot{\pi}\right)\right)}^{3}+8\left(H_a H_b+H_a H_c+H_b H_c\right) G_{4XXX} {\left(\left(\dot{\pi}\right)\right)}^{4},                             \nonumber  &
\end{flalign}
\begin{flalign}
  &A_{15}^{i} = -F_{X}+K_{\pi}-K_{X \pi} {\left(\left(\dot{\pi}\right)\right)}^{2}-2K_{X X} \left(\ddot{\pi}\right) {\left(\left(\dot{\pi}\right)\right)}^{2}-2K_{X} \left(\ddot{\pi}\right)&\nonumber\\
  &-2\left(H_j+H_k\right) K_{X} \left(\dot{\pi}\right)+6G_{4X \pi} \left(\ddot{\pi}\right)+6\left(H_j + H_k\right) G_{4X \pi} \left(\dot{\pi}\right)&\nonumber\\
  &+4G_{4 X X \pi } \left(\ddot{\pi}\right) {\left(\left(\dot{\pi}\right)\right)}^{2}-4\left(H_j+H_k\right) G_{4 X X \pi } {\left(\left(\dot{\pi}\right)\right)}^{3}-8\left(H_j+H_k\right) G_{4XXX} \left(\ddot{\pi}\right) {\left(\left(\dot{\pi}\right)\right)}^{3}&\\
  &-12\left(H_j+H_k\right) G_{4XX} \left(\dot{\pi}\right) \left(\ddot{\pi}\right)-4\left({H_j}^{2} + \left(\dot{H_j}\right)+{H_k}^{2}+\left(\dot{H_k}\right)+ 3 H_j H_k\right) G_{4XX} {\left(\left(\dot{\pi}\right)\right)}^{2}&\nonumber\\
  &+2G_{4 X \pi \pi} {\left(\left(\dot{\pi}\right)\right)}^{2}-2\left(\left(\dot{H_j}\right)+{H_j}^{2} + \left(\dot{H_k}\right) + {H_k}^{2} + H_j H_k\right) G_{4X},\nonumber&
\end{flalign}
\begin{flalign}
 &{A_{17}} = F_{\pi}-2F_{X \pi} {\left(\left(\dot{\pi}\right)\right)}^{2}-8\left(H_b H_c+H_a H_c+H_a H_b\right) G_{4X \pi} {\left(\left(\dot{\pi}\right)\right)}^{2}& \nonumber\\
 &+2\left(H_a+H_b+H_c\right) G_{4\pi\pi} \left(\dot{\pi}\right)-8\left(H_a H_b+H_a H_c+H_b H_c\right) G_{4 X X \pi } {\left(\left(\dot{\pi}\right)\right)}^{4}&\nonumber\\
 &+4\left(H_a+H_b+H_c\right) G_{4 X \pi \pi} {\left(\left(\dot{\pi}\right)\right)}^{3}+K_{\pi \pi} {\left(\left(\dot{\pi}\right)\right)}^{2}-2\left(H_a+H_b+H_c\right) K_{X \pi} {\left(\left(\dot{\pi}\right)\right)}^{3}&\\
 & +2\left(H_a H_c+H_b H_c+H_a H_b\right) G_{4\pi},&\nonumber
\end{flalign}
\begin{flalign}
  &A_{18}^{i} = -2F_{X} \left(\dot{\pi}\right)-2K_{X \pi} {\left(\left(\dot{\pi}\right)\right)}^{3}+2K_{\pi} \left(\dot{\pi}\right)-4K_{X X} \left(\ddot{\pi}\right) {\left(\left(\dot{\pi}\right)\right)}^{3}-4K_{X} \left(\dot{\pi}\right) \left(\ddot{\pi}\right)&\nonumber\\
  &-4\left(H_a+H_b+H_c\right) K_{X} {\left(\left(\dot{\pi}\right)\right)}^{2}-4\left(\left(\dot{H_j}\right) +\left(\dot{H_k}\right) + \left(H_j + H_k\right)^2 \right.+&\nonumber\\
  &\left. 2 H_b H_c+2 H_a H_b+2H_a H_c\right) G_{4X} \left(\dot{\pi}\right)+4G_{4 X \pi \pi} {\left(\left(\dot{\pi}\right)\right)}^{3}+8G_{4 X X \pi } \left(\ddot{\pi}\right) {\left(\left(\dot{\pi}\right)\right)}^{3}&\\
  &+12G_{4X \pi} \left(\dot{\pi}\right) \left(\ddot{\pi}\right) +8\left(H_a + H_b+H_c\right) G_{4X \pi} {\left(\left(\dot{\pi}\right)\right)}^{2}+2\left(H_j+H_k+2H_i\right) G_{4\pi}&\nonumber\\
  &-8\left( \left(H_j + H_k\right)^2 +\left(\dot{H_j}\right)+\left(\dot{H_k}\right)+2H_i H_j+2H_i H_k + H_j H_k\right) G_{4XX} {\left(\left(\dot{\pi}\right)\right)}^{3}&\nonumber\\
  &-8\left(H_j+H_k\right) G_{4 X X \pi } {\left(\left(\dot{\pi}\right)\right)}^{4}-16\left(H_j+H_k\right) G_{4XXX} \left(\ddot{\pi}\right) {\left(\left(\dot{\pi}\right)\right)}^{4}&\nonumber\\
  &-32\left(H_j+H_k\right) G_{4XX} \left(\ddot{\pi}\right) {\left(\left(\dot{\pi}\right)\right)}^{2}-4\left(H_j + H_k\right) G_{4X} \left(\ddot{\pi}\right),\nonumber&
\end{flalign}
\begin{flalign}
  & {A_{20}} = \frac{1}{2}F_{\pi \pi}-F_{X \pi \pi} {\left(\left(\dot{\pi}\right)\right)}^{2}-2F_{X X \pi} \left(\ddot{\pi}\right) {\left(\left(\dot{\pi}\right)\right)}^{2}-F_{X \pi} \left(\ddot{\pi}\right)                                 \nonumber& \\
  & -\left(H_a+H_b+H_c\right) F_{X \pi} \left(\dot{\pi}\right)+K_{\pi \pi} \left(\ddot{\pi}\right)+\left(H_a+H_b+H_c\right) K_{\pi \pi} \left(\dot{\pi}\right)+K_{X \pi \pi} \left(\ddot{\pi}\right) {\left(\left(\dot{\pi}\right)\right)}^{2} \nonumber& \\
  & -\left(H_a+H_b+H_c\right) K_{X \pi \pi} {\left(\left(\dot{\pi}\right)\right)}^{3}-2\left(H_a+H_b+H_c\right) K_{XX \pi} \left(\ddot{\pi}\right) {\left(\left(\dot{\pi}\right)\right)}^{3}                                                   \nonumber& \\
  & -\left( \left(H_a + H_b + H_c\right)^2 +\left(\dot{H_a}\right)+\left(\dot{H_b}\right)+\left(\dot{H_c}\right) \right) K_{X \pi} {\left(\left(\dot{\pi}\right)\right)}^{2}                                                                  \nonumber & \\
  & -2\left(H_a+H_b+H_c\right) K_{X \pi} \left(\dot{\pi}\right) \left(\ddot{\pi}\right)+\frac{1}{2}K_{\pi \pi \pi} {\left(\left(\dot{\pi}\right)\right)}^{2}-2\left( \dfrac{d}{dt} \left[H_a H_b+H_a H_c+H_b H_c\right] \right.               \nonumber & \\
  & \left. +{H_a}^{2} \left(H_b + H_c\right) + H_b^2 \left(H_a + H_c\right) + H_c^2 \left(H_a + H_b\right) + 3H_a H_b H_c\right) G_{4X \pi} \left(\dot{\pi}\right)                                                                            \nonumber & \\
  & +6 \left(H_a+H_b+H_c\right) G_{4 X \pi \pi} \left(\dot{\pi}\right) \left(\ddot{\pi}\right)+2\left( \left(H_a + H_b + H_c\right)^2 + \left(\dot{H_a}\right) \right.                                                                         & \\
  & \left. +\left(\dot{H_b}\right)+\left(\dot{H_c}\right)\right) G_{4 X \pi \pi} {\left(\left(\dot{\pi}\right)\right)}^{2} +4\left(H_a+H_b+H_c\right) G_{4 XX \pi \pi} \left(\ddot{\pi}\right) {\left(\left(\dot{\pi}\right)\right)}^{3}       \nonumber& \\
  & -4\left(H_a H_b+H_a H_c+H_b H_c\right) G_{4 XX \pi \pi} {\left(\left(\dot{\pi}\right)\right)}^{4}-8\left(H_a H_b+H_a H_c+H_b H_c\right) G_{4 XXX \pi} \left(\ddot{\pi}\right) {\left(\left(\dot{\pi}\right)\right)}^{4}                    \nonumber& \\
  & -16\left(H_a H_b+H_a H_c+H_b H_c\right) G_{4 X X \pi } \left(\ddot{\pi}\right) {\left(\left(\dot{\pi}\right)\right)}^{2}-4\left(  \dfrac{d}{dt} \left[H_a H_b+H_a H_c+H_b H_c\right] \right.                                               \nonumber& \\
  & \left. +{H_a}^{2} \left(H_b + H_c\right) + H_b \left(H_a + H_c\right) + H_c^2 \left(H_a + H_b\right) + 3H_a H_b H_c\right) G_{4 X X \pi } {\left(\left(\dot{\pi}\right)\right)}^{3}                                                        \nonumber& \\
  & +2\left(H_a+H_b+H_c\right) G_{4 X \pi \pi \pi} {\left(\left(\dot{\pi}\right)\right)}^{3}-2\left(H_a H_b+H_a H_c+H_b H_c\right) G_{4X \pi} \left(\ddot{\pi}\right)                                                                         \nonumber & \\
  & +\left(\left(\dot{H_c}\right)+{H_c}^{2}+\left(\dot{H_b}\right)+{H_b}^{2}+H_b H_c+\left(\dot{H_a}\right)+{H_a}^{2}+2H_a H_c+H_a H_b\right) G_{4\pi\pi},                                                                                    \nonumber&
\end{flalign}
\begin{flalign}
  &B^{ab} = \frac13 (H_b - H_a) A_1,&\\
  &B^{bc} = \frac13 (H_c - H_b) A_1,&\\
  &B^{ac} = \frac13 (H_c - H_a) A_1.&
\end{flalign}
\section*{Appendix B}
\begin{align}
  &D_1 = \frac23\left( H_a - H_b \right) \bar{H}_a \bar{H}_b+\frac23\left(H_c - H_b\right) \bar{H}_b \bar{H}_c+ \frac23 \left( H_a - H_c \right) \bar{H}_a \bar{H}_c+&\\
  &+\frac{1}{3}\sum_{\substack{i,j = a,b,c\\ i\neq j}}\frac{d}{dt}\left[\bar{H}_i \bar{H_j}\right],&\\
  &D_2 = \frac{1}{a b c} \frac{d}{dt} \left[a b c D_1\right],&
\end{align}

\begin{align}
  &A = \frac23 {A_{1}} \left(\bar{H}_a \bar{H}_b+\bar{H}_a \bar{H}_c + \bar{H}_b \bar{H}_c\right) - \frac{9}{2} \frac{{{A_{4}}}^{2}}{{A_{3}}}.&
\end{align}
In this section we denote
\[A_4 = \frac13 \sum_{l = a,b,c} A_4^l \bar{H}_l, \quad A_8 = \frac13 \sum_{l = a,b,c} A_8^l \bar{H}_l, \quad A_{18} =  \sum_{l = a,b,c} A_{18}^l \bar{H}_l,\]

and as usual $i \neq j \neq k$.
\begin{flalign}
  &\Theta_i = \frac{1}{3}{A_{1}} \left(\bar{H}_j+\bar{H}_k\right),&
\end{flalign}
\begin{flalign}
  &\Lambda_i = \frac{1}{3 A_3}\left(9{A_{4}} A_{8}^{i}+{A_{1}} {A_{11}} \left(\bar{H}_j+\bar{H}_k\right)\right),&
\end{flalign}
\begin{flalign}
  &\Xi_i = \frac12 \left[-2A_{13}^{ij} \bar{H}_j-2A_{13}^{ik} \bar{H}_k-4\eta A_{2}^{jk} \bar{H}_j \bar{H}_k-\eta \frac{d}{dt} \left[\frac{{A_{1}} {A_{4}} \left(\bar{H}_i+\bar{H}_k\right)}{{A_{3}}}\right] -\right.&\nonumber\\
  & \frac{1}{3} \frac{d}{dt}\left[\frac{{A_{1}} {A_{11}} \left(\bar{H}_i+\bar{H}_k\right)}{{A_{3}}}\right]+\frac{1}{3 A_3}\left(-3A_{8}^{i}+2{A_{1}} \eta \left(\bar{H}_j+\bar{H}_k\right)\right) \sum_{l = a,b,c} A_4^l \dot{\bar{H_l}} +&\nonumber\\
  &+\left.\frac{1}{3}\frac{A_{1}}{A_3} \left({A_{11}}+3\eta {A_{4}}\right) \left(\bar{H}_j+\bar{H}_k\right) H \left(\bar{H}_i-\bar{H}_j-\bar{H}_k\right) +\frac{1}{3 A_3}{A_{1}} {A_{17}} \left(\bar{H}_j+\bar{H}_k\right) \right],&
\end{flalign}
\begin{flalign}
  &\Pi_i = \frac12 \left[\frac{1}{2} \frac{d}{dt} \left[\frac{{A_{1}} {A_{4}} \left(\bar{H}_j+\bar{H}_k\right)}{{A_{3}}}\right] - \frac{1}{3}\frac{A_{1}}{{A_{3}}} \left(\bar{H}_j+\bar{H}_k\right) \sum_{l = a,b,c} A_4^l \dot{\bar{H_l}} +\right.&\nonumber\\
  &\left. +\frac{1}{2}\frac{{A_{1}} {A_{4}} H}{{A_{3}}} \left(\bar{H}_j+\bar{H}_k\right) \left(\bar{H}_j+\bar{H}_k-\bar{H}_i\right)+2A_{2}^{jk} \bar{H}_j \bar{H}_k \right], &
\end{flalign}
\begin{flalign}
  &\Sigma_i = \frac{1}{3}{A_{1}} \eta \left(\bar{H}_j+\bar{H}_k\right)-A_{8}^{i},&
\end{flalign}
\begin{flalign}
  &M =   - \frac{1}{2} \frac{1}{a b c} \frac{d}{dt}\left[a b c \left( - \frac{1}{2}\frac{{A_{11}} {A_{17}}}{{A_{3}}} + \eta \left(\dot{\eta}\right) A+\eta \left(-{A_{18}}+2\sum_{l=a,b,c} A_{8}^{l} \dot{\bar{H}_l}-3\dot{{A_{8}}} -\right.\right.\right.&\nonumber\\
  &\left.\left.\left. - \frac{3}{2} \frac{{A_{17}}{A_{4}}}{A_3}-9{A_{8}} H+\frac12 \frac{{A_{11}}}{{A_{3}}} \sum_{l=a,b,c} A_{4}^{l} \dot{\bar{H}}_l \right)\right)\right]  - \frac{1}{4}\frac{{{A_{17}}}^{2}}{A_{3}} + {A_{20}}+\frac{1}{2} A {\left(\dot{\eta}\right)}^{2} + &\nonumber\\
  & +\left(\dot{\eta}\right) \left(-{A_{18}}+2\sum_{l=a,b,c} A_{8}^{l} \dot{\bar{H}_l}-3\dot{{A_{8}}} - \frac{3}{2} \frac{{A_{17}} {A_{4}}}{A_3}-9{A_{8}} H+\frac12 \frac{{A_{11}}}{{A_{3}}} \sum_{l=a,b,c} A_{4}^{l} \dot{\bar{H}_l} \right) + &\nonumber\\
  &+\eta \left(-\frac12 \frac{A_{17}}{A_3} \sum_{l=a,b,c} A_{4}^{l} \dot{\bar{H}_l} + \frac12 \frac{1}{abc} \frac{d}{dt}\left[\frac{a b c A_{11}}{A_3} \sum_{l=a,b,c} A_{4}^{l}\dot{\bar{H}_l}\right]\right)+&\nonumber\\
  &+{\eta}^{2} \left(\frac{1}{2}{A_{1}} {D_{2}}+\frac{1}{2}{D_{1}} \left(\dot{{A_{1}}}\right)- \frac14 \frac{1}{{A_{3}}} \left(\sum_{l=a,b,c} A_{4}^{l}\dot{\bar{H}_l}\right)^2 + \frac34 \frac{1}{abc} \frac{d}{dt}\left[\frac{abc A_4}{A_3}\sum_{l=a,b,c} A_{4}^{l}\dot{\bar{H}_l}\right]\right),&
\end{flalign}
\begin{flalign}
  & m = \left({A_{1}} {D_{2}}+{D_{1}} \dot{{A_{1}}}\right) - \frac{1}{A_3} \left(\sum_{l=a,b,c} A_{4}^{l}\dot{\bar{H}_l}\right)^2 + \frac{1}{abc} \frac{d}{dt} \left[\frac{a b c A_4}{A_3} \sum_{l=a,b,c} A_{4}^{l}\dot{\bar{H}_l}\right].&
\end{flalign}